\documentclass[11pt]{article}
\pdfoutput=1

\usepackage{graphicx}

\usepackage{multicol} 

\usepackage{color}
\usepackage{latexsym}

\definecolor{rosso}{cmyk}{0,1,1,0.4}
\definecolor{rossos}{cmyk}{0,1,1,0.55}
\definecolor{rossoc}{cmyk}{0,0.5,1,0.2}
\definecolor{blu}{cmyk}{1,1,0,0.3}
\definecolor{blus}{cmyk}{1,1,0,0.6}
\definecolor{blucc}{cmyk}{1,0.4,0.2,0}
\definecolor{viola}{cmyk}{0,1,0,0.6}
\definecolor{viola2}{cmyk}{0,1,0.2,0.6}
\definecolor{verde}{cmyk}{0.92,0,0.59,0.25}
\definecolor{verdec}{cmyk}{0.92,0,0.59,0.15}
\definecolor{verdes}{cmyk}{0.92,0,0.59,0.4}
\font\tenrsfs=rsfs10 at 12pt
\font\sevenrsfs=rsfs7
\font\fiversfs=rsfs5
\newfam\rsfsfam

\textfont\rsfsfam=\tenrsfs
\scriptfont\rsfsfam=\sevenrsfs
\scriptscriptfont\rsfsfam=\fiversfs
\def\mathscr#1{{\fam\rsfsfam\relax#1}}

\def\circa#1{\,\raise.3ex\hbox{$#1$\kern-.75em\lower1ex\hbox{$\sim$}}\,}

\usepackage{amsmath,latexsym,amssymb,color,hyperref,graphicx}
\newcommand{\be}{\begin{equation}}
\newcommand{\ee}{\end{equation}}
\newcommand{\bea}{\begin{eqnarray}}
\newcommand{\ena}{\end{eqnarray}}
\newcommand{\no}{\noindent}
\newcommand{\nb}{\nonumber}

\renewcommand\o{\omega}  
\renewcommand\a{\alpha}
\renewcommand\b{\beta}

\newcommand\m{\ensuremath{\mu}}
\renewcommand\k{\ensuremath{\kappa}}

\newcommand\n{\ensuremath{\nu}}

\newcommand\tr{\text{tr}}

\newcommand{\de}{\partial}
\newcommand{\ha}{\frac{1}{2}}

\newcommand{\ba}{\begin{eqnarray}}
\newcommand{\ea}{\end{eqnarray}}

\makeatletter
\def\ps@mine{%
    \def\@oddfoot{\hfil\thepage\hfil}\let\@evenfoot\@oddfoot
    \let\@oddhead\@evenhead%
    \let\@mkboth\@gobbletwo
    \let\sectionmark\@gobble
    \let\subsectionmark\@gobble
    }
\renewcommand\section{\@startsection {section}{1}{\z@}%
                                   {-3.5ex \@plus -1ex \@minus -.2ex}%
                                   {2ex \@plus.2ex}%
                                   {\normalfont\large\sffamily\bfseries}}
\renewcommand\subsection{\@startsection {subsection}{1}{\z@}%
                                   {-3.5ex \@plus -1ex \@minus -.2ex}%
                                   {2ex \@plus.2ex}%
                                   {\normalfont\sffamily\bfseries}}
\makeatother

\begin{document}

\def\FILL{\hfill\hfill\hfill}

\title{\sffamily\bfseries 
Cosmology of bigravity with doubly  coupled matter   \\[3ex]
  \normalsize
   D. Comelli$^a$, M. Crisostomi$^b$, K. Koyama$^b$, L. Pilo$^{c,d}$ and G. Tasinato$^{e}$\\[2ex]
   \it\small
   \hspace{2cm}$^a$INFN, Sezione di Ferrara,  I-35131 Ferrara, Italy\FILL\\
   \hspace{2cm}$^b$Institute of Cosmology and Gravitation, University of Portsmouth, Portsmouth, PO1 3FX, UK\FILL\\
   \hspace{2cm}$^c$Dipartimento di Fisica, Universit\`a di L'Aquila,  I-67010 L'Aquila, Italy\FILL\\
   \hspace{2cm}$^d$INFN, Laboratori Nazionali del Gran Sasso, I-67010 Assergi, Italy\FILL\\
   \hspace{2cm}$^e$Department of Physics, Swansea University, Swansea, SA2 8PP, UK\FILL\\[2ex]
   \hspace{2cm}{\tt comelli@fe.infn.it}, {\tt marco.crisostomi@port.ac.uk}, {\tt kazuya.koyama@port.ac.uk},\FILL\\
   \hspace{2cm}{\tt luigi.pilo@aquila.infn.it}, {\tt g.tasinato@swansea.ac.uk}\FILL\\[-6ex]
}

\date{\small \today}

\maketitle

\thispagestyle{empty}


\def\abstractname{\sc Abstract}
\begin{abstract}
\no
We study  cosmology in  the bigravity formulation of the dRGT model where matter couples to both metrics.
At linear order in perturbation theory two mass scales emerge: an hard one from the dRGT potential,  and an  environmental dependent one from the coupling of bigravity with matter.
At early time, the dynamics is dictated by the second mass scale which
is of order of the Hubble scale.
The set of gauge invariant perturbations that  couples to matter follow  closely
the same behaviour as in GR.
The remaining perturbations show no issue in the scalar sector, while
problems arise in the tensor and vector sectors.
During radiation domination, a tensor mode grows power-like at  super-horizon scales.  More dangerously, 
the only propagating vector mode features an exponential instability on sub-horizon scales. We  discuss
 the consequences of such instabilities and 
speculate on possible ways to deal with them. 
\end{abstract}

\bigskip

\section{Introduction and Summary}
The physical mechanism responsible for the present day acceleration of
our universe is unknown. The simplest explanation is a positive
cosmological constant; however the large amount of tuning required to
fit the data seems excessive. Alternatives based on modifications of
Einstein General Relativity (GR), which become observationally
relevant at large scales, are being actively explored nowadays
\cite{megarevs}. Among them, Massive Gravity~\cite{deRham} has received
special attention: from an effective field theory perspective, it is
one of the most natural options to investigate when renouncing to the
diffeomorphism invariance of GR. In such a scenario, the graviton mass introduces a new energy scale that can be related with the scale of dark energy.  In its simplest incarnation,
to build a massive deformation of GR a reference non-dynamical metric 
is  needed. Besides  the  aether-like nature of the reference metric, an   unattractive feature
from a theoretical perspective,   
there are various   motivations to go beyond  massive gravity  and enter  in the realm of
bigravity theories~\cite{Damour,Berezhiani,Comelli,Hassan:2011zd}. For example,
spatially flat homogenous
Friedmann-Robertson-Walker (FRW) solutions  do not
exist~\cite{FRWfroz} in Lorentz invariant ghost free massive gravity,  and even allowing for open FRW solutions~\cite{open}
strong coupling~\cite{tasinato} and ghostlike
instabilities~\cite{defelice} develop. Flat FRW solutions exist~\cite{Comelli:2013tja,lang} in the
case of Lorentz breaking models~\cite{Rubakov,dub,uscan,uslong,Comelli:2014xga}.

In bigravity one can find various branches  of regular cosmological
solutions describing flat FRW cosmologies~\cite{cosm-CCNP,bi-GF}. A
branch where the gravity modification is  equivalent to an effective
cosmological constant suffers of strong coupling~\cite{cosmpert}.
A more promising branch is unstable at early time,
being characterized by an exponential growth of fluctuations
\cite{cosmpert,pertslast}. Also, singular FRW-type
solutions exist~\cite{cosm-CCNP} which exhibit only  mild instabilities (power-law growth of vector and tensor modes \cite{Lagos:2014lca,Cusin:2014psa}). 
On the other hand, they correspond to bouncing universes characterized by a naked curvature singularity, which makes their physical relevance questionable. Recently, it has been proposed to extend the theory of massive (bi)gravity by considering a more general coupling to matter -- called doubly matter coupling -- in which the physical metric coupled to the matter energy momentum tensor is an appropriate linear combination of the 
two metrics \cite{deRham:2014naa, Heisenberg:2014rka} (see also \cite{Tamanini:2013xia, Noller:2014sta} for different approaches). This scenario, although problematic for massive gravity \cite{Solomon:2014iwa}, is potentially
interesting in the bigravity setup since a qualitatively new 
branch of FRW cosmological solutions exists \cite{Enanderbef}, hence
its cosmological perturbations deserve to be investigated.
Such a theory is not ghost-free~\cite{deRham-ghost, Yamashita:2014fga}, but
there exist  physically interesting situations where the
Boulware-Deser (BD) ghost does not  represent 
an immediate phenomenological problem. This is the case if its
  mass is above the 
cut-off scale $\Lambda_c$ of  the theory under consideration.  In
addition,  $\Lambda_c$  might be parametrically larger than
the strong coupling scale where the effects of the  graviton mass term
become important and interesting. Moreover, the ghost does 
not manifest itself at linear order in an expansion in fluctuations
around particularly symmetric configurations (as for example FRW
cosmologies). In this work, after studying the two branches of cosmological solutions at the  
homogeneous level, we focus on  the dynamics of linearized
cosmological fluctuations around the new  
homogenous backgrounds allowed by the doubly matter coupling. No
hints of BD ghost mode are found at linear level in fluctuations, and
the theory propagates the seven degrees of freedom as expected
for a healthy bigravity theory. The dynamics of scalar  fluctuations is healthy, and no instabilities are found in this sector. The dynamics of tensor and vector fluctuations is  richer, but it shows problematic behaviours. The tensor sector exhibits a power-like instability at superhorizon scales during the radiation domination era. We argue that such instability is not 
extremely serious, and  can be tamed by an appropriate choice of
initial conditions, possibly  motivated by inflation.   
Much worse is the behaviour of vector fluctuations. In this case, during
radiation domination,  we find a gradient instability at subhorizon scales, which leads to an exponential growth of small scale fluctuations, rapidly  driving  the theory outside the regime of validity of perturbation theory.  Hence, this serious instability rules out the 
cosmological configurations that we consider. 
Nevertheless, we speculate on possible  extension of the bigravity theory under consideration, that might be able to cure such instability problems.    

\section{The theory under consideration}
\subsection{Scalar Field}
\label{scalarfield}
Before plunging in the study of bigravity, it is interesting to understand in a simplified setting 
the peculiar feature of the non-minimal coupling of matter to gravity
proposed by Ref.~\cite{deRham:2014naa}. 
We will show how the consistency of the effective description of matter as a (perfect) fluid
necessarily requires the dynamical character of the second metric,
selecting bigravity as the only consistent formulation.
Take as matter a scalar field  $\phi$ that couples
with gravity not simply by the metric $g_{\mu \nu}$ but trough a
combination  of $g_{\mu \nu}$ and a non-dynamical flat metric $f_{\mu \nu}$
\be
ds_f^2 =f_{\mu \nu} dx^\mu dx^\nu = -{z'}^2 \, dt^2+ \, dx^i dx^j
\delta_{ij} = \de_\mu \Phi^a \de_\nu \Phi^b  \eta_{ab}\, dx^\mu dx^\nu
\, .
\label{aux}
\ee
We denoted by $'$ the time derivative with respect to $t$.
In order to restore diffeomorphism  (diff) invariance, the non-dynamical metric $f$ can be written using four Stuckelberg
fields; 
in the unitary gauge we have  $\Phi^0 = z(t) $ and $\Phi^i = x^j
\, \delta^i_{\,\,j} $.
The scalar field $\phi$, with a potential $F(\phi)$, couples to gravity according to 
\be
S= \int d^4 x \left[2 \, M_{pl}^2 \,\sqrt{g}\, R -\sqrt{G} \left(\ha\, G^{\mu \nu} \de_\mu \phi\, \de_\nu
\phi + F(\phi) \right) \right]  \, .
\label{anscal}
\ee
Thus, $\phi$ is minimally coupled to $G_{\mu\nu} $  defined by
\be
G_{\m\n}= \a^2 g_{\m\n} + 2\,\a\,\b\, g_{\m\rho}Y{}^\rho_\n + \b^2 f_{\m\n} \,,
\label{effmet}
\ee
where $Y{}^\mu_\nu=(\sqrt{X})^\mu_\nu$ 
and $X{}^\mu_\nu = g^{\mu \sigma} {f}_{\sigma \nu}$ and $\a\,,\b$ are
two arbitrary constants. Notice that $\phi$ is not minimally
  coupled to $g_{\mu \nu}$. Setting $\beta=0$
and $\alpha=1$, we recover the standard minimal coupling. 
It is clear from the ADM canonical  analysis that the total Hamiltonian
is not anymore linear in the lapse $N$ and shift $N^i$ of the
dynamical metric $g_{\mu \nu}$, then the number of propagating degrees of
freedom (DoF) will be more than 3. The seemingly innocent action
(\ref{anscal}) actually represents a modification of gravity. 

Let us consider FRW homogeneous cosmological solutions where 
\be
ds^2 = -N^2(t) \, dt^2 + a^2(t) \, dx^i dx^j \delta_{ij} \,  ,
\ee
and 
\be
\begin{split}
&ds^2_{\text{eff}}= G_{\mu \nu} dx^\mu dx^\nu = 
- N^2_{\text{eff}} \, dt^2 + a_{\text{eff}}^2 \, dx^i dx^j \delta_{ij} \, ; \\
& N_{\text{eff}} = \alpha \, N + \beta \, z' \, , \qquad \qquad a_{\text{eff}} =
\alpha \, a +\beta \, . 
\end{split} 
\ee
The Energy Momentum Tensor (EMT) for the scalar is diagonal,
fluid-like and can be written as
\be
\begin{split}
& T^0_0= \rho_\phi  \, , \qquad \qquad T^i_j= p_\phi\, \delta^i_j \,,\\ 
&  \rho_\phi = \alpha\, \frac{a_{\text{eff}}^3}{a^3} \left( \frac{\phi'^2}{2
    \, N_{\text{eff}}^2} + V \right) \, , \;\;
      p_\phi = \alpha \,\frac{N_{\text{eff}} \,
  a_{\text{eff}}^2}{N\, a^2} \left( \frac{\phi'^2}{2
    \, N_{\text{eff}}^2} - V \right) \, ;
\end{split}
\label{MEMT}
\ee
which has a
peculiar dependence on $N$ and $z'$.
The expression for the EMT reduces to the standard one when $\beta \to 0,
\, \alpha \to 1$.

Contrary to the case of a scalar field minimally coupled to gravity,
the time-time and the spatial components of the Einstein equations and the equation of motion for $\phi$ are
all independent. Indeed, taking the time derivative of
the time-time component of the Einstein 
equations and using the equation of motion of $\phi$, one can solve for
$a''$; then inserting this expression in 
the spatial components of the Einstein equations one gets the following
constraint\footnote{We do not consider unphysical cases where $z'$ and/or $a_{\text{eff}}=0$.}
\be
\beta \, p_\phi =0 \,.
\label{conb1}
\ee
Thus, unless $\beta =0$, Einstein equations
require that $p_\phi=0$. The same constraint follows from the
requirement that the scalar EMT is conserved. Of course such constraint
has no counterpart in GR, where the EMT for $\phi$ is automatically
conserved when $\phi$ satisfies its equation of motion.

Hence:
\begin{itemize}
\item[-]
in a FRW background the dynamics of a scalar field in (\ref{anscal}) is not equivalent to
a perfect fluid;
\item[-]
the pressure has to vanish.
\end{itemize}
Such difficulties were not taken into account in~\cite{Gumrukcuoglu:2014xba}.

In the following we will show that both issues are absent  when
the non-dynamical Stuckelberg metric~(\ref{aux}) is promoted to
a full-fledged dynamical one, see also~\cite{Solomon:2014iwa}. 
The reason for such a behaviour can be traced back to the
non-dynamical nature of the metric $f_{\mu \nu}$ used in the new
matter coupling. The conservation of the energy momentum tensor
\be
\nabla^{(g)}_\nu T^{\mu \nu} =0 \, ,
\ee
defined as the response of the matter action $S_\text{matt}$ to a
diffeomorphism variation of the dynamical metric
\be
\delta S_\text{matt} = -\ha \int d^4 x \sqrt{g} \, T^{\mu \nu} \, \delta
g_{\mu \nu} \, ,
\ee
gives such a strong constraint that in general cannot be satisfied
unless very special conditions like (\ref{conb1}) are met. This is not
very surprising and it is typical of theories with non-dynamical object~\cite{strau,giulini}.

When instead the metric $f_{\m\n}$ gets dynamical, inserting its own Ricci scalar in the action,
for the following FRW parametrization
\be
ds_f^2 = -{z'}^2 \, dt^2 + \o^2(t) \, dx^i dx^j \delta_{ij} \,  ,
\ee
condition (\ref{conb1}) becomes
\be
\beta \left(z'\,a' - N\, \o' \right) p_\phi=0 \,,
\label{conb2}
\ee
and a new possibility of realizing (\ref{conb2}) opens up.
Within this new way, the scalar field dynamics can be still captured
by the perfect fluid description and no spurious constraint is required.

Therefore, for the rest of this paper, we will parametrize
the matter content of the Universe 
through a perfect fluid and the metric $f_{\mu \nu}$ entering in
(\ref{effmet}) will become dynamical 
in the bigravity formulation.

\subsection{Bigravity and Matter Coupling}
Consider the action of massive bigravity with the dRGT potential as
interaction between the 
two dynamical metrics $g_{\mu \nu}$ and $f_{\mu \nu}$
\be
S=
\int d^4 x  \left\{ \,\sqrt{g} \left[ 
M_{pl}^2 \left( {\cal R}
-2  m^2   \, V \right) \right] + \sqrt{f} \,   \kappa   \, M_{pl}^2\;
\widetilde {\cal R} \,\right\}+ 
S_{\text{matt}} \, .
\label{act} 
\ee
In the presence of two metrics, it is not a priori clear to what metric
matter couples to. 
In general, the BD ghost revives in the
presence of doubly coupled 
matter~\cite{deRham-ghost, Yamashita:2014fga}. In \cite{deRham:2014naa}
a new matter coupling was proposed 
where matter is minimally coupled to the effective metric $G_{\m\n}$
given by (\ref{effmet}).
%
Although the BD ghost persists even with this special doubly coupled
bigravity model, it was shown that the BD ghost 
does not appear in the decoupling limit \cite{deRham:2014naa,deRham-ghost}. 
As for the scalar field non-minimally coupled to gravity of section
\ref{scalarfield}, the theory described by the action (\ref{act})
propagate more than the seven DoF expected in the bigravity formulation of
dRGT. The extra scalar mode is most probably a ghost, however, the
theory is still acceptable if the mass of such a mode is above the
ultraviolet cutoff $\Lambda_c$. Clearly this point deserves further investigation.

The matter EMT is defined as the response of the matter action to a
variation of $g\,(f)$
\be
\delta S_{\text{matt}} = - \frac{1}{2} \int d^4x \sqrt{G} \, {\cal T}^{\mu \nu}
\delta G_{\mu \nu} = - \frac{1}{2} \int d^4x \left ( \sqrt{g} \,
  T^{\m \n} \delta g_{\m \n} +\sqrt{f} \,
  \widetilde{T}^{\m \n} \delta g_{\m \n} \right) \, .
\ee
The modified Einstein equations can be written as
\begin{gather}
\label{eqm1}
\,{E}^\mu_ \nu  +  Q{}^\m_\n = \frac{1}{2\;M_{pl}^2 }\, T^\mu_\nu \,,  \\
\label{eqm2}
\kappa  \, {\widetilde E}^\mu_\nu  +  \widetilde Q{}^\m_\n =
\frac{1}{2\;M_{pl}^2 }\, 
\widetilde{T}^\mu_\nu \,,
\end{gather}
where $Q$ ($\widetilde Q$)  are the effective energy-momentum tensors
induced by the  interaction term for the two metrics.

Let us introduce the BD ghost free potential~\cite{Gabadadze:2011, Hassan:2011vm, HR}
\be
\label{eq:genpot}
\qquad V=\sum_{n=0}^4 \, a_n\, V_n \,, 
\vspace*{-2ex}
\ee 
\vskip.3cm
\no
where the $V_n$ are the symmetric polynomials of $Y$
%
\be
\begin{split}
&V_0=1\,,\qquad 
V_1=\tau_1\,,\qquad
V_2=\tau_1^2-\tau_2\,,\qquad
V_3=\tau_1^3-3\,\tau_1\,\tau_2+2\,\tau_3\,,\\[1ex]
&V_4=\tau_1^4-6\,\tau_1^2\,\tau_2+8\,\tau_1\,\tau_3+3\,\tau_2^2-6\, \tau_4\,,
\end{split}
\ee 
with $\tau_n=\tr(Y^n)$.  We have that 
\bea
 Q{}_\nu^\mu &=&  { m^2}\, \left[ \;  V\; \delta^\mu_\nu \,  - \,  (V'\;Y)^\mu_\nu  \right] \,, \\[1ex]
 \widetilde Q{}_\nu^\mu &=&  m^2\, q^{-1/2} \, \; (V'\;Y)^\mu_\nu \,,
\ena
where  $(V^\prime)^\mu_\nu = \de V / \de Y_\mu^\nu$ and  $q =\det
X=\det(f)/\det(g)$.

\section{Homogeneous cosmological solutions}
\subsection{FRW Ansatz and Conservation Laws}
Let us consider homogeneous FRW bi-diagonal metrics with flat spatial slices in conformal time, so
the form of $g$ and $f$ is as follows:
\be
\begin{split}
ds^2 &=   a^2(\tau) \left(- d\tau^2 + dr^2 + r^2 \, d \Omega^2 \right) \\
\widetilde{ds}^2 &= \omega^2(\tau) \left[- c^2(\tau) \, d\tau^2 + dr^2+ r^2 \, d \Omega^2 \right] \, 
\label{frw}
\end{split}
\ee
The effective metric gets the following form
\be
ds^2_{\text{eff}}= - (\alpha \, a + \beta\, c\, \o )^2 d\tau^2 + (\alpha \, a +
\beta \, \omega)^2 \left( dr^2 + r^2\, d\Omega^2 \right) \, .
\label{geff}
\ee
Consistency of the equations of motion requires the following
Bianchi-type constraints
\be
\begin{split}
& \nabla_\mu \left(2 Q^\mu_\nu - M_{pl}^{-2} \, T^\mu_\nu \right) =0 \,
, \\
& \tilde{\nabla}_\mu \left(2 \tilde{Q}^\mu_\nu - M_{pl}^{-2} \,
  \tilde{T}^\mu_\nu \right) =0 \, .
\end{split}
\ee
When ${\cal T}_{\mu \nu}$ is the EMT of a perfect fluid, i.e.  ${\cal T}_{\mu \nu}=(p+\rho)\;u_\mu\;u_\nu+p\;G_{\mu\nu}$ with $u_\alpha\,G^{\alpha\beta}\,u_\beta=-1$, the previous
equations can be combined to give
\bea
&& 3 (w+1) \left(\alpha\,  a\, {\cal H}+\beta\,  \omega\, {\cal H}_\omega
   \right) \rho+ \left(\alpha\, a+\beta\,  \omega\right)\rho' = 0 \,, \label{newmattcons} \\[1.5ex]
&& \left(c\, {\cal H} - {\cal H}_\o \right) \left[w\,\a\,\b \left(\alpha\, 
   a+\beta\,  \omega\right)^2 \rho -2\, m^2 M_{pl}^2 \left(a_1\, a^2+4\,
   a_2\, a\, \omega +6\, a_3\, \omega^2\right)\right] =0 \, , 
\label{bcons}
\ena
where $
{\cal H}= a'/a $ and ${\cal H}_\o=
\omega'/\omega$ and the equation of state $w$,  $p = w \, \rho$.
Notice that condition (\ref{newmattcons}) corresponds to the conservation of the matter EMT ${\cal T}^{\mu \nu}$
with respect to the metric $G_{\m\n}$.
The constraint (\ref{bcons}) can be realised in two inequivalent
ways:
\vskip .3cm
\no
{\bf Branch 1}
\vskip .3cm

\no
In this branch eq (\ref{bcons}) is realised through the vanishing of the square bracket, i.e. the pressure of the fluid is determined by the massive potential.
Notice that at early time, when $\rho \gg m^2 \, M_{\text{pl}}^2 $,
consistence requires that $w \, \rho \approx 0$, even though the Universe
should be radiation dominated at that epoch. In this branch 
we have to deals with the very same issue found in section
\ref{scalarfield}
and the description of matter as a perfect fluid is not consistent with the one of a 
scalar field.
This branch was studied
in~\cite{Gumrukcuoglu:2014xba}, when  $f_{\mu
  \nu}$ is a flat non-dynamical metric, in presence of a scalar field. Though, contrary to the case of massive gravity with minimally coupled
matter, flat FRW solutions exist, the non-physical requirement of
$w=0$ makes the present branch not very appealing as discussed in Ref.~\cite{Solomon:2014iwa}. 


\vskip .3cm
\no
{\bf Branch 2}
\vskip .3cm

\no
In this case
\be
c= \frac{{\cal H}_\o}{\cal H} \,. 
\label{solc}
\ee
Notice that the limit $\beta \to 0$ exists and we recover the very same branch of
FRW cosmology in bigravity with standard matter minimally coupled to
the metric $g_{\mu \nu}$. Contrary to branch 1, since condition
(\ref{solc}) is matter independent, the scalar field dynamics is
equivalent to a perfect fluid; in this sense matter has the standard
effective description. 
It is also interesting to note that (\ref{solc}) is equivalent to
requiring that the matter's action and the interaction part of the action
are separately diff invariant, namely 
 \be
\begin{split}
& \nabla_\nu  Q^{\mu \nu} =0 \, \quad f \text{ is on-shell} \, , \\
& \nabla_\nu T^{\mu\nu} = 0 \, \quad f \text{ and } \phi \text{ are on-shell}  \,;
\end{split}
\ee
and equivalently for the metric $f$.
Clearly, the branch two is the most interesting one and  from now on we will
focus on it.

\subsection{Background solutions for Branch 2}
Introducing the ratio of the two scale factors $\xi= \o/a$,
the $tt$-component of the modified Einstein equations for $g$ reads
\be
\frac{{\cal H}^2}{a^2} = \frac{1}{6 M_{pl}^2} \rho  \left(\alpha +\beta
\xi\right)^3 + m^2 \left(a_0/3+a_1\, \xi+2\, a_2\, \xi^2+2\,
a_3\, \xi^3\right) \,.
\label{hubble}
\ee
The relation between $\xi$ and $a$ is
determined by using (\ref{solc}) and (\ref{hubble})  in the
$tt$-component of the modified Einstein equations  for $f$; the result is
an algebraic equation 
\be
\begin{split}
& 2 m^2 \left[a_1 + \left(6\, a_2-\kappa\,a_0 \right) \xi +3
   \left(6\, a_3 - \kappa\,a_1\right) \xi ^2 + 6 \left(4\, a_4 -\kappa\,a_2 \right) \xi^3 -6\, \kappa\,a_3\, \xi ^4 \right] \\
&= M_{pl}^{-2} \rho  \left(\alpha +\beta  \, \xi \right)^3 \left(\kappa\,\alpha \, \xi -\beta \right) \, .
\end{split}
\label{constr}
\ee
The case $\a \to 0$ and $\b \to
0$ has been already studied in \cite{cosmpert} and instabilities were
found in the scalar sector.

Throughout this paper we assume that the scale of the graviton mass
$m$ is of the order of the present 
Hubble scale, i.e $\; m^2 \sim M_{pl}^{-2} \, \rho_\Lambda \sim
H_0^2$. Such a choice is the most natural one if massive gravity has
anything to do with the present acceleration of the Universe.
Then, according to the evolution equation for the matter energy density (\ref{newmattcons}) that gives
\be
\rho=\rho_0 \left[\left(\alpha + \beta \, \xi \right) a\, \right]^{-3(1+w)} \,,
\ee
at early times, provided that $\xi \ll 1/a$, we can always consider $m^2 M_{pl}^{2}/ \rho \ll1$ as a dimensionless expansion parameter.
Therefore,  at early time, solutions of equation (\ref{constr}) can be
classified according to the different regimes of 
$\xi$, i.e $\xi \ll1,\, \xi \gg1$, and $\xi \sim 1$.

\begin{itemize}

\item[$\blacktriangleright$] $\xi \ll1$

In this case the solution for the matter energy density is given by
\be
\begin{split}
\rho &= -2 \, m^2 \, M_{pl}^{2} \, 
\frac{ a_1 }{\beta\,\alpha^3} \\
& + \frac{  2 \, m^2 \, M_{pl}^{2} \, 
   \xi \left[\kappa\,a_0\, \alpha \, \beta
   -a_1\left( \kappa\,\alpha^2 -3\, \beta^2\right) - 6\,a_2\, \alpha
   \, \beta \right]}{\alpha^4 \beta ^2} + 
{\cal O}(\xi^2) \, .
\end{split}
\ee
At the leading order we have $\rho \sim \rho_\Lambda$. This rules out
the small $\xi$ regime at  early times.

\item[$\blacktriangleright$] $\xi \gg1$

In this case we have that at early times the matter energy density is
given by
\be
\begin{split}
\rho &= -12 \, m^2 \, M_{pl}^{2}\, \frac{a_3\, }{\a\,\b^3} \\
& -  
  \frac{12 \, m^2 \, M_{pl}^{2} \left[\kappa\, a_2\, \alpha \, \beta
   -a_3\left(3\, \kappa\,\alpha^2 - \beta^2\right) - 4\,a_4\, \alpha
   \, \beta \right]}{\kappa\, \alpha ^2 \beta ^4 \, \xi}\, 
+ {\cal O}(\xi^{-2}) \, .
\end{split}
\ee
The behaviour is equivalent to a cosmological constant; thus, also the
large $\xi$ regime is not suitable for early time cosmology.

\item[$\blacktriangleright$] $\xi \simeq 1$

For this last case we have two solutions for $\xi$, and at the leading order they read
\be
\xi = \begin{cases}
-\frac{\a}{\b} + {\cal O}\left(\frac{m^2M_{pl}^{2}}{\rho}\right) \,,\\[1ex]
\,\,\,\frac{\b}{\kappa\,\a } + {\cal O}\left(\frac{m^2M_{pl}^{2}}{\rho}\right) \,.
\end{cases}
\ee
When $\xi \simeq -\frac{\a}{\b}$ the spatial components of the
effective metric (\ref{geff}) are singular and moreover the early time
cosmology is dominated by a cosmological constant, i.e. ${\cal H}^2 \propto m^2\,a^2$.
When instead $\xi \simeq \frac{\b}{\kappa\,\a }$, at the leading order we have
\be
c = 1\,, \qquad\qquad {\cal H}^2 =\frac{ a^2}{6\, M_{pl}^2} 
\frac{\left( \kappa\,\a^2 + \b^2\right)^3}
{\kappa^3\,\a^2}\,\rho \,,
\label{newfreed}
\ee
and the corrections are of the order $\sim {\cal O}(m^2 M_{pl}^{2}/ \rho)$.
Therefore, up to a renormalisation of the Newtonian constant\footnote{We
should compare the coefficient in front of $a^2\,\rho$ in (\ref{newfreed}) with the one appearing in spherically symmetric solutions through the Vainshtein mechanism \cite{vain} using this new matter coupling.}, the
early time cosmology 
 is very
similar to GR at the background level. 

The dynamics of the background is described by the following equations 
\bea 
\frac{d \xi}{d \ln a} &=& (c-1) \xi \,,  \nonumber\\
\frac{d \rho}{d \ln a} &=& - 3 \left( 1 + \frac{\beta(c-1) \xi}{\alpha + \beta \xi}\right) 
(1+w) \rho \,,
\label{dynamical}
\ena
where the first equation follows from the definition of $\xi$ and the second equation is the continuity equation for matter. The lapse function in the second metric, $c$, can be found combining the Einstein equations and can be expressed in terms of $\xi$. The late time fixed point is given by $c=1$ and $\rho=0$, which corresponds to a de Sitter phase. At early times, $\xi \simeq \frac{\b}{\kappa\,\a }$ and $c=1$. Once the density becomes lower, the solutions are attracted towards the de Sitter fixed point. This transition happens when $m^2 M_{pl}^2/\rho \sim 1$. At the de Sitter fixed point we have $H^2 \sim m^2$, this is why, in order to explain the late time acceleration of the Universe, we need to assume $m \sim H_0$.  Fig.~1 shows an example of the two dimensional phase plain spanned by $\rho$ and $\xi$ and the trajectory of the background solution that connects the early time cosmology to the de Sitter fixed point. 

\begin{figure}[ht]
  \centering{
  \includegraphics[width=3in]{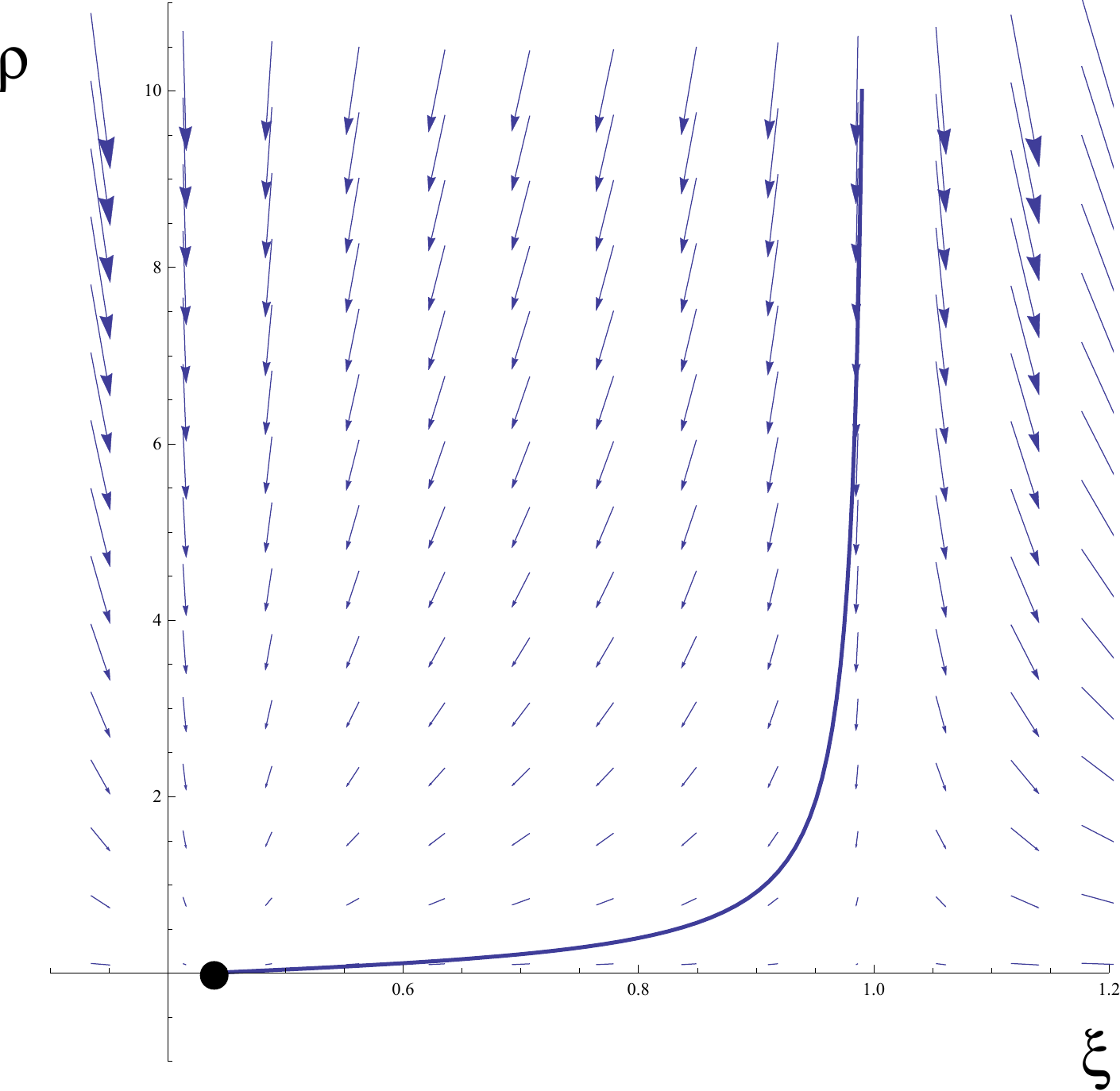}
  }
  \caption{The phase plane described by eq (\ref{dynamical}). We take $a_0=a_1=1,\, a_2=a_3=a_4=0,\,\alpha=\beta=1,\,\kappa=1,\,w=1/3$ as an illustration.}   
\end{figure}

Notice that the $\xi \sim 1$ regime is incompatible with the $\beta \to 0$ limit.\footnote{We stress that in the discussion of the various regimes we have considered $\beta \neq0$.}
Indeed, when $\beta$ is very small, $\xi \sim {\cal O}\left(m^2M_{pl}^{2}/\rho\right)$ and 
we turn back to the small $\xi$ regime studied in~\cite{cosmpert}. 
Thus the early time cosmology for $\xi \simeq \frac{\b}{\kappa\,\a }$ exists only when $\beta \gg m^2 M_{pl}^2/\rho$. In order to have this early time cosmology until the solution reaches the de Sitter point, $\beta$ needs to be ${\cal O}(1)$, assuming that all other parameters are also ${\cal O}(1)$.  

In the next section we will analyse cosmological perturbations around this background.

\end{itemize}

\section{Cosmological Perturbations}
Perturbations around a FRW background can be classified according to
representations of the $SO(3)$ group, namely scalar, vector and tensor
modes. It is convenient to use the gauge invariant formulation
following~\cite{cosmpert,pertslast} and for the benefit of the reader the
relevant expression are collected in Appendix~\ref{pert}. 

In order to avoid cluttering in the main text, the general expressions
of the perturbed Einstein equations for the scalar and vector sector
are given in Appendices~\ref{scal} and \ref{vect}, see also \cite{Schmidt-May:2014xla} for a different derivation.  

The general feature that emerges is that comparing with the ghost free
bigravity theory with minimally coupled matter~\cite{cosmpert}, the
new matter coupling gives rise to an additional {\it environmental dependent
mass}; namely, the new equations can be obtained from the old ones by
the replacement
\be
m^2 \to m^2 +  J(a\,, c,  \, \xi) \,  \frac{\rho}{M_{pl}^2} \, , 
\label{envmass}
\ee
where $J$ is a function of time that depends on the field under
consideration. 
The bottom line is that two mass scales are present: a
hard mass~$m^2$ coming from the 
interaction potential for the metrics, see~(\ref{act});
and, as a consequence of the non-minimal coupling, a
second ``soft'' running mass scale.
In the physical branch, where $\xi \sim 1$, at early times the soft mass always dominates on the hard one, since  $\rho/M_{pl}^2 \gg m^2$.
Thus at early time the potential $V$ plays no role and can be
simply neglected at the leading order in $\sim {\cal O}\left(m^2M_{pl}^{2}/\rho\right)$. Of course this is the case only when $\beta \neq
0$, otherwise the regime $\xi \sim 1$ simply does not exist and the
$m^2$ corrections become the leading part.
The perturbed equations of motion can be fully solved in this approximation.

Notice that the new matter coupling is driven by the matter EMT ${\cal
T}^{\mu \nu}$, so the contribution to the $ij$ component of the
perturbed Einstein equations is always proportional to $w$; as a
result such a contribution vanishes when the Universe is dominated by
non-relativistic matter where $w=0$.

In the following we will often use the Fourier transform  with respect
to $x^i$ of the various 
perturbations, the corresponding
3-momentum will be denoted $k^i$ and $k^2 = k^i k^i \delta_{ij}$. To keep notations as
simple as possible we will use the same name for the field and its Fourier transform.

\subsection{Scalar Sector}
In the scalar sector, see Appendix~\ref{scal}, the fields  ${\cal E}, \, {\cal B}_1$
and $\Psi_{1/2}$ are non dynamical and can be expressed in
terms of $\Phi_{1/2}$, basically one of the Bardeen potentials for $g$
and $f$, that satisfy two second order equations; thus  2 scalar DoF
propagate. As already pointed out  in section~\ref{scalarfield}, by
using canonical analysis an additional scalar is expected, however as matter of
fact such a mode does not propagate around a homogeneous background. Nevertheless it is expected to appear in a less symmetric background and/or
at higher order in perturbation theory~\cite{deRham-ghost}. The
consequences of the missing 
BD   mode, at leading order in  cosmological perturbations around FRW cosmologies, 
 deserve 
further study.

In order to solve for the two propagating fields, it is convenient to use the following combination of themselves:
\be
\Phi_+ \equiv \Phi_1 + \frac{\b^2}{\k\,\a^2} \Phi_2 \,, \qquad \Phi_- \equiv \Phi_1 - \frac{\b^2}{\k\,\a^2} \Phi_2 \,.
\ee
It is $\Phi_+$ that couples to matter thus $\Phi_+$ is the relevant metric perturbation for observations. 
$\Phi_+$ turns out to decouple completely (at the leading order) and satisfies
the same equation as in  GR. Instead the equation of motion 
for $\Phi_-$ is more complicated; at early times it can be expanded
in powers of  $\tau/\tau_U \ll1$, where  $\tau_U$  is the age of the
universe in conformal time. In what follows, we shall consider the
leading order in such an expansion.

Consider first the case of a radiation dominated Universe.

For $\Phi_+$ we have the very same equation of GR, i.e.
\be
\Phi _+'' +\frac{4 }{\tau }\, \Phi _+'+ \frac{k^2}{3}\, \Phi _++{\cal
  O}\left(\frac{m^2M_{pl}^{2}}{\rho} \right) =0 \, .
\ee
 For $\Phi_-$, outside the horizon where  $k\,\tau \ll 1$ we get 
\be
\Phi _-'' +\frac{6}{\tau}\, \Phi _-'+ \frac{5}{\tau^2}\,\Phi _-  - \frac{\left(\k\,\a^2-\b^2 \right)}{\left(\k\,\a^2+\b^2\right)}\left( \frac{2}{\tau}\, \Phi _+' +\frac{5}{\tau^2}\,\Phi _+ \right)+{\cal O}\left(\frac{\tau}{\tau_U} \right)=0 \,,
\ee
and for the modes inside the horizon, namely $k\,\tau \gg 1$, we have
\be
\Phi _-'' +\frac{7}{\tau}\, \Phi _-'+ \frac{97}{12\,\tau^2}\,\Phi _-  - 
\frac{\left(\k\,\a^2-\b^2 \right)}{\left(\k\,\a^2+\b^2\right)}
\left(\frac{3}{\tau}\, \Phi _+' - \frac{k^2}{3}\,\Phi _+\right) +{\cal O}\left(\frac{\tau}{\tau_U} \right)=0 \,.
\ee
Both scalars show no instability and their non-decaying modes are constant outside the horizon.
Inside the horizon instead they both oscillate with a different decaying amplitude and frequency.
The corresponding energy density perturbation is at the leading order
\be
\frac{\delta \rho_{\text{gi}}}{\rho}= 
\frac{2\,\k\,\a^2}{3\left(\k\,\a^2+\b^2 \right)}\left( 3+ k^2\,\tau^2\right) \Phi_+ \,,
\label{drsurrad}
\ee
and therefore behaves like in GR.
Outside the horizon it turns out to be constant, instead inside the horizon it has a fixed amplitude oscillating behaviour. 
In the case of matter dominated Universe both the fields $\Phi_+$ and
$\Phi_-$ obey the same equation of GR at 
the leading order, namely
\be
\Phi _{+/-}'' +\frac{6}{\tau }\, \Phi _{+/-}'+{\cal O}\left(\frac{\tau}{\tau_U} \right)
=0 \, .
\ee
It is worth to stress  that the gauge-invariant density
  perturbation $\delta \rho_{\text{gi}}/\rho$ is
the real observable quantity and is given by
\be
\frac{\delta \rho_{\text{gi}}}{\rho}= 
\frac{\k\,\a^2}{6\left(\k\,\a^2+\b^2 \right)}\left( 12+ k^2\,\tau^2\right) \Phi_+ \,.
\label{drsurmat}
\ee
Again the behaviour of $\delta \rho_{\text{gi}}/\rho$ is similar to
GR: outside the horizon it is frozen and inside the horizon it
grows like $\tau^2$ leading to the formation of structures.

In order to study the other Bardeen potentials, it is useful to introduce also the combinations
\be
\Psi_+ \equiv \Psi_1 + \frac{\b^2}{\k\,\a^2} \Psi_2 \,, \qquad \Psi_- \equiv \Psi_1 - \frac{\b^2}{\k\,\a^2} \Psi_2 \,.
\ee
In general it turns out that
\be
\Psi_+ + \Phi_+ = 0 \,, \qquad \Psi_- + \Phi_- = - \frac{w\,\rho\,a^2 \,\b^2 \left( \k\,\a^2+\b^2\right)^2}{\k^3\,\a^2\,M_{pl}^{2}}\, {\cal E} \,.
\ee
It is interesting to note that in bigravity the double diff
invariance is broken down to the diagonal diff invariance when an
interaction between the two metric is introduced. The $+$ sector
is  protected by diagonal diffs and indeed in the
tensor sector corresponds to massless spin 2 modes, while the $-$ gives rise
to the massive ones, see also \cite{Schmidt-May:2014xla}.

Remarkably, the instability that was present in the field $\Phi_2$ for
modes inside the horizon with matter 
minimally coupled to $g_{\m\n}$, found in~\cite{cosmpert}, is not
present. For such a behaviour the new matter coupling is
instrumental, indeed the $\xi~\sim~1$ regime for the background
emerges and the very fast gradient instabilities disappear. However,
as we will see soon, instabilities in the tensor and vector sectors emerge.

\subsection{Tensor Sector}

The first hint of the problematic behaviour of fluctuations manifests itself when studying the tensor fluctuations.
For this sector, the final expression for
the perturbed Einstein equations is rather simple
\bea
&&{h^{TT}}_{1 \, ij}'' + 2 {\cal H} \, {h^{TT}}_{1 \, ij}' - \Delta
{h^{TT}}_{1 \, ij} + a^2 \left[m^2\,f_1- 
w\,\bar \rho\, d\,\xi \right] \left(h^{TT}_{1 \, ij} -h^{TT}_{2 \, ij} \right) =0 \,,
 \label{1t} \\[.3cm]
 && {h^{TT}}_{2 \, ij}'' +\left[2 \left(  {\cal H} +\frac{ \xi'}{\xi} \right)
 -   \,\frac{ c'}{c} \right] {h^{TT}}_{2 \, ij}'  - c^2   \, \Delta
{h^{TT}}_{2 \, ij} \nb \\
 && \qquad\qquad\qquad\qquad\quad + \frac{a^2\,c}{\kappa\, \xi^2} \left[ m^2 \,f_1 - w\,\bar \rho\, d\,\xi \right]
   \left(h^{TT}_{2 \, ij} -h^{TT}_{1 \, ij} \right) =0 \, ,
\label{2t}
\ena
where $f_1,\,\bar \rho$ and $d$ are defined in Appendix \ref{scal}.
Also in this sector, it is useful to introduce the following combination of fields
\be
h_+ \equiv h_1 + \frac{\b^2}{\k\,\a^2} h_2 \,, \qquad h_- \equiv h_1 - \frac{\b^2}{\k\,\a^2} h_2 \,,
\label{tcomb}
\ee
with the indices and the ${}^{TT}$ symbol understood.
Again it is $h_+$ that is relevant for observed gravitational waves, since it is the combination appearing inside $G_{\m\n}$.

In particular, for the physical background solution, we get for general  $w$
\bea
&&h_+''+2\, {\cal H} \,h_+'+k^2\,h_+ = 0 \, ; \label{2ten1}  \\[0.2cm]
&&h_-''+2\, {\cal H} \,h_-' + \left[ k^2 + a^2 \left(\k\,\a^2+\b^2\right)\left( \frac{m^2\,f_1}{\b^2}- \frac{w\left(\k\,\a^2+\b^2\right)^2\bar\rho}{\k^3\,\a^2}\right)\right]\,h_- \nb \\
&& \qquad - a^2 \left(\k\,\a^2-\b^2\right) \left[ \frac{m^2\,f_1}{\b^2}- \frac{w\left(\k\,\a^2+\b^2\right)^2\bar\rho}{\k^3\,\a^2}\right]\,h_+ =0 \, . \label{2ten2}
\ena 
The combination $h_+$ behaves as in GR and represents spin 2 masseless
modes protected by diagonal diff invariance.  Instead the combination $h_-$
features a contribution to its mass 
proportional to $w$ (see the parenthesis
 in the first line of eq \eqref{2ten2}). This contribution is entirely due 
 to  the new matter coupling.  If $w$ is positive, the
 mode  $h_-$ is tachyonic. If $w$ is negative, this  mode acquires  a
 positive mass squared.
  
Let us discuss some physical implications of the evolution equation \eqref{2ten2}  for  
the mode $h_-$. During matter domination, $w=0$ and the new contributions to the
effective mass  to  the mode $h_-$ vanish: there are no instabilities associated with the new
coupling of gravity with matter. 
   
In the radiation dominated era, $w=1/3$: the mode $h_-$ acquires a tachyonic mass, and
an instability is expected.  Indeed, radiation dominated
super-horizon solutions for $h_+$ and $h_-$, in terms of the
scale factor $a\propto \tau$, are
\bea
h_+&=&  \left( C_1 - \frac{C_2}{a} \right) \,,
\\
h_- &=& \frac{\left( \k\,\a^2 - \b^2 \right)}{\left( \k\,\a^2 + \b^2 \right)} \left( C_1 - \frac{C_2}{a} \right) + C_3\,a^{-\frac{\sqrt{5}+1}{2}}+C_4\,a^{\frac{\sqrt{5}-1}{2}} \,, \label{solt2}
\ena
 where $C_1,\,C_2,\,C_3,\,C_4$ are integration constants. The quantities  $C_1,C_2$
control the healthy evolution of the mode $h_{+}$, and contribute also to the mode $h_-$
 due to the source term in the second line of eq \eqref{2ten2}. The integration constants $C_2$, $C_3$ 
  correspond to 
   decaying modes, that can be neglected, while $C_4$ controls a growing mode.  
  As a consequence, this system features a mild power-like instability at superhorizon scales, but only 
    during radiation dominated era. 
    
   Hence, 
   in order to ensure that this  set-up is  under perturbative control, we need to impose that the 
   amplitude of $h_-$ does not exceed unity during radiation domination. 
  Let us choose  units  in which 
   the scale factor after the reheating period, the beginning of radiation domination, is
    $a_{rh}=1$.   
   Focus on the contribution
   of the growing mode only, and take into account the evolution of the scale factor during radiation domination. We find   that at the end of the radiation period, at matter-radiation equality, 
    we have to satisfy the inequality
   \be
C_4\,\left(a_{eq}\right)^{\frac{\sqrt{5}-1}{2}}\,=\,
   C_4\,\left(\frac{T_{rh}}{T_{eq}} \right)^{\frac{\sqrt{5}-1}{2}}\,<\,1
   \ee
  where the temperature of transition from radiation to matter domination is $T_{eq}\,\simeq\,3$ eV, while
  the reheating temperature when the  radiation era starts is model dependent, 
 and depends on the reheating temperature after inflation. 
   Taking a representative  value
  $T_{rh}\,=\,10^{8}$ GeV,  we find an upper bound for the integration 
constant $C_4$:

\be
C_4\,<\,10^{-10}\,. \label{ine1}
\ee  

Interestingly, 
 such small values
for the  quantity $C_4$, 
 can be motivated by 
the inflationary phase that precedes radiation domination. Indeed, it is inflation
that sets the initial condition for the amplitude of tensor fluctuations, that evolve in the radiation 
dominated era. 
We sketch here an argument to explain
this fact.  
Inflation is a phase of quasi-de Sitter expansion, where the parameter $w\simeq-1$. During this
phase, the mode $h_{-}$ acquires a positive mass squared  in its
evolution equation \eqref{2ten2} due to the new coupling. The solution of the coupled system of equations
 \eqref{2ten1}, \eqref{2ten2}  is then (the scale factor scales as $a\,\simeq\,1/(- H \tau)$)
 \bea
h_+&=&  \left( D_1 - \frac{D_2}{a^3} \right) \,,
\\
h_- &=& \frac{\left( \k\,\a^2 - \b^2 \right)}{\left( \k\,\a^2 + \b^2 \right)} \left( D_1 - \frac{D_2}{a^3} \right) + D_3\,a^{-\frac{3}{2}} \,\sin{\left(\frac{\sqrt{3}}{2}\,\ln{a} \right)}+D_4\,\,a^{-\frac{3}{2}} \,\cos{\left(\frac{\sqrt{3}}{2}\,\ln{a} \right)} \,, \label{teninfl2}
\ena
 with $D_1\dots D_4$
integration constants.   
The quantities $D_1$, $D_2$ control the evolution of the mode
   $h_+$:
    neglecting the decaying mode $D_2$, the constant mode $h_{+}$  matches continuously
    between the inflationary  and radiation dominated era, hence we  set $D_1=C_1$. 
      The quantities $D_3$, $D_4$
   control the specific properties of the mode $h_-$, so  they  determine the initial conditions
   for the mode  $C_4$ at the beginning
   of radiation era, after the end of inflation. 
    Notice that $D_3$, $D_4$   are   decaying modes. 
    We can write the matching relation between
   solutions \eqref{solt2} and  \eqref{teninfl2}
   \be
   D_4\,\simeq\,C_4, \label{valC4}
   \ee
   where we make the crude assumption of  a sudden transition from inflation to radiation domination, so that
   the end of inflation occurs at $a_{rh}=1$. 
     Assuming that during inflation the size of $h_-$ remains bounded, and that there are no cancelations among the different terms  in eq \eqref{teninfl2}, at the beginning of inflation we have 
  the condition
  \be
  D_{4}\,a_{in}^{-3/2}\,\le\,1\,. \label{conD4}
  \ee  
 Notice that in our units in which $a_{rh}=1$, the value of $a_{in}$, the scale factor at the beginning of
  inflation, is very small. So in these units it is  natural to choose a very small value for $D_4$
  to satisfy condition \eqref{conD4}. 
   Indeed,
saturating the previous equality, and using the relation $a_{rh}/a_{in}\,=\,e^{N_{ef}}$, with $N_{ef}$ the
e-fold number (that we take for definiteness $N_{ef}\,=\,60$), recalling that in our units
$a_{rh}=1$, we find
\be
D_4\,=\, D_{4}\,a_{rh}^{-3/2}\,=\,D_{4}\,a_{in}^{-3/2}\,e^{-3 N_{ef}/2}\,\simeq\,10^{-39}\hskip1cm\Rightarrow 
\hskip1cm
D_4\,\simeq\,C_4\simeq10^{-39}\,,
\ee
that comfortably satisfies  inequality \eqref{ine1}.

\smallskip

Of course the previous arguments are based on linear perturbation theory.
   Non-linear effects couple different modes:   higher order  fluctuations in  the scalar  sector are expected to feed the initial value of the tensor amplitude at the beginning of radiation era. 
    On the other hand, in estimating their  effects, one would have 
to take in careful consideration the coupling factors between scalars and $h_-$, that can be suppressed by factors of the graviton mass. So it is possible that even taking into account non-linearities, the 
growth of tensor fluctuations can still be maintained under control: it would be interesting
to investigate more in detail this topic.  

\subsection{Vector Sector}
The dynamics of fluctuations in the vector sector is unfortunately much more problematic. 
 Using the conservation of the matter EMT ${\cal T}^{\mu \nu}$ with
 respect to the effective metric $G_{\m\n}$ 
(equation (\ref{vecc}) in Appendix \ref{pert}), the velocity
perturbations $\delta v_i$ can be completely solved.
From equations (\ref{1vts}-\ref{2vss}) in Appendix \ref{vect}, one can show that all vectors can
be expressed in terms of a single combination ${\cal V}_{12}$
that satisfies a second order equation; thus, only a single transverse
vector propagates. 
 As for the tensors, the qualitative features of vector dynamics depends on the equation of state
 of the fluid constituting the matter EMT. When the fluid equation of state $p\,=\,w \,\rho$ is such that $w\le 0$, no
 instabilities are found and the system is healthy. Instead, we find serious gradient instabilities when
 $w>0$. 
 In particular, in the case of a radiation dominated universe we
have that $\delta v = \delta v_0$,  where $\delta v_{0}$ is an arbitrary constant, and at the leading order in~$\tau/\tau_U$~\footnote{Spatial indices
for the vectors are understood.}
\bea
&&V_1 =-\frac{8}{k^2\,\tau^2}\,\delta v_0 - \frac{5\,\b^2}{\left(\k\,\a^2+\b^2\right)\left(k^2\,\tau^2+5\right)}\,{\cal
  V}_{12}' \, ;\\[0.2cm]
&&V_2= -\frac{8}{k^2\,\tau^2}\,\delta v_0 + \frac{5\,\k\,\a^2}{\left(\k\,\a^2+\b^2\right)\left(k^2\,\tau^2+5\right)}\,{\cal
  V}_{12}' \, ;\\[0.2cm]
&&{\cal V}_{12}'' +\frac{10}{\tau\left(k^2\,\tau^2+5\right)}\,{\cal
  V}_{12}'
-\frac{\left(k^2\,\tau^2+5\right)}{5\,\tau^2}\,{\cal V}_{12}
=0 \, .\label{eqv12}
\ena
So we find the `wrong sign' in front of the gradient term for the propagating mode ${\cal V}_{12}$, leading to a gradient instability in the sub-horizon limit $k \tau \gg\,1$. In this regime, we have an exponential growth for
 ${\cal V}_{12}$, and the leading contribution to the solution of \eqref{eqv12} is
 \be
 {\cal V}_{12} \,\propto\,e^{\frac{k\,\tau}{\sqrt{5}}}\,.
 \ee
Also when selecting tuned initial conditions, such exponential instability is so severe that even a small
initial amplitude of vector generated by non-linear effects (as the ones mentioned at the end of the previous
section)  will be rapidly amplified to a level that drives
perturbation theory out of control. 
Notice that once more in the $+$ sector no instability is
present. Indeed, the combination
\be
V_+ = V_1 + \frac{\b^2}{\kappa \, \a^2} \, V_2 =
-\frac{8}{k^2\,\tau^2} \left( 1+ \frac{\b^2}{\kappa \,
    \a^2} \right) \delta v_0 \, ,
\label{vcomb}
\ee
represents a decreasing mode. Thus instabilities are present only in the
$-$ sector. 
To conclude, the dynamics of vector fluctuations ruin the cosmology of
bigravity doubly coupled to matter. Possible 
ways-out, to be investigated in the future, could be to modify our ansatz for the EMT.
 Indeed in  this work we mainly  focussed on an EMT of perfect fluid
 form: it might be that other choices of matter 
content can lead to better behaved system. As an example, one could further include  vector degrees of freedom, also  non-minimally coupled to  gravity (as recently explored in massive gravity or related systems in~\cite{deRham:2014tga,Tasinato:2014eka,Heisenberg:2014rta}), 
and study cosmological configurations in the scenario  of doubly coupled matter.  
 Acting as additional source, they might be able  to fix the problems we found  with vector fluctuations.  
We leave this interesting issues to a future investigation. 

\section{Conclusions}
Massive gravity has been the subject of an extensive investigation. The
main phenomenological motivation is to explain the present
acceleration of the universe. However devising a satisfactory model is
not an easy task. The simplest ghost free Lorentz invariant version
of massive gravity~\cite{Gabadadze:2011} with an auxiliary
non-dynamical metric has no flat homogenous solution~\cite{FRWfroz} and only Lorentz breaking~\footnote{The term refers to the existence (non existence) of Lorentz global symmetry in the gravitational sector in the unitary gauge.} models~\cite{Rubakov,dub,uscan,uslong,Comelli:2014xga} can
support such configurations~\cite{Comelli:2013tja}. Sticking with Lorentz
invariant models, to overcome the  limitations of a non-dynamical   auxiliary
metric, one can promote it to a dynamical one in the contest of
bigravity. As far as homogeneous configurations are concerned, in  bigravity theories 
the situation drastically improves with respect to massive gravity,  and one finds branches of
flat FRW solutions. Leaving aside the ones with a curvature
singularity at late time with $c <0$ (see eq. (\ref{frw})) -- as discussed
in~\cite{cosm-CCNP} and more recently
in~\cite{Konnig:2014xva,Lagos:2014lca,Cusin:2014psa} -- regular background solutions
unfortunately features an exponential instability in the scalar
sector at early time, quickly invalidating perturbation theory
already during radiation domination \cite{cosmpert}. Things do not change when both
metrics are coupled with two different matter
sectors~\cite{pertslast}.
In the presence of two metrics there is a certain degree of ambiguity
on how to couple matter with gravity~\cite{deRham:2014naa} and actually
one can consider a sort of democratic coupling of the two metrics with
matter, see eq. (\ref{effmet}). Though the new coupling reintroduce the Boulware-Deser ghost~\cite{deRham-ghost}, one can argue that its mass is above the cutoff and does not affect the low energy physics in the spirit of effective field theory. One of the interesting features of the new
coupling is that it gives rise to an effective background dependent soft
mass, see for instance  (\ref{envmass}) for the case of an FRW
background. As a result even taking the ``hard'' mass $m$ in the
deforming potential to zero, we still have a massive gravity theory
thanks to the environmental soft mass, that for an FRW background is
proportional to the Hubble parameter. This is the feature that opens up a new dynamical regime compared with minimally coupled case such that $\xi = \omega/a$ stays constant at early time and then flows toward a de Sitter attractor responsible for the present acceleration phase, see figure 1. The next step is to study the behaviour of cosmological perturbations. Things get better in the scalar sector where no instability is found, however troubles develop in the tensor and specially in the vector sector. Among the tensor modes the $+$ combination (see eq.(\ref{tcomb})) which is protected by diagonal diffs has the same dynamics as in GR, while the second independent tensor mode develops a power-law growth until the matter domination. Such a growth is naturally counterbalanced by a sufficient low primordial production during inflation. In the vector sector the situation is more worrisome: while again the $+$ combination of vector fields, see eq.(\ref{vcomb}), has only a decreasing mode, the  $-$ combination shows an exponential instability when $w>0$ at subhorizon scales. The main effect is to  loose  theoretical control of the theory at the perturbative level already during the radiation domination. On the other hand, we stress that in all cases studied the anomalous growth is only present in sectors that do not couple directly with observed matter;
 moreover, we  speculated on possible  extension of the bigravity theory under consideration, that might be able to cure such instability problems.    

\subsection*{Acknowledgments}
MC would like to thank Tony Padilla and David Stefanyszyn for many interesting discussions, beers and their kind hospitality.
KK is supported by the UK Science and Technology Facilities Council grants number ST/K00090/1 and ST/L005573/1. 
GT is supported by an STFC Advanced Fellowship ST/H005498/1. 

\begin{appendix}

\section{Perturbed geometry}
\label{pert}
Let us now consider the perturbations of the FRW background (\ref{frw})
\be
g_{\mu \nu} = \,  \bar g_{\mu \nu}  +a^2 \,  {h_1}_{\mu
  \nu} \, , \qquad f_{\mu \nu} =  \bar f_{\mu \nu}  +
 \o^2 \,  h_{2 \, \mu \nu}  \,,
\ee
parametrized as follows 
\be
\begin{split}
& {h}_{1 \, 00} \equiv - 2 A_1  \, , \qquad  {h}_{2 \, 00}  \equiv - 2 c^2 \, A_2 \,, \\
& {h}_{1/2 \, 0 i}  \equiv {\cal C}_{1/2 \,i} - \de_i B_{1/2}  \, , \qquad
\de^i {\cal V}_{1/2 \, i}=  \de^i {\cal C}_{1/2 \, i} = \de^j {h^{TT}}_{1/2 \, ij} =
\delta^{ij} {h^{TT}}_{1/2 \, ij}  =0  \, ,\\
& h_{1/2 \, ij}  \equiv {h^{TT}}_{1/2 \, ij} + \de_i {\cal V}_{1/2 \, j} + \de_j {\cal
  V}_{1/2\, i} + 2 \de_i \de_j E_{1/2} + 2 \, \delta_{ij} \, F_{1/2}  \,.
\end{split}
\ee
Spatial indices are raised/lowered using the spatial flat metric.

Under a gauge transformation generated by $\zeta^\mu$ the metric perturbation transforms
\be
\begin{split}
& \delta h_{1 \, \mu \nu } = a^{-2} \left(\zeta^\alpha \de_\alpha  {\bar g}_{
\mu \nu} +  \bar g_{\alpha \nu}  \, \de_\mu
  \zeta^\alpha + \bar g_{\mu \alpha}  \, \de_\nu \zeta^\alpha \right)
\, , \\[.2cm]
&\delta h_{2 \, \mu \nu } = \o^{-2} \left(\zeta^\alpha \de_\alpha \, {\bar
  f}_{\mu \nu } +  \bar f_{\alpha \nu}  \, \de_\mu
  \zeta^\alpha + \bar f_{\mu \alpha}  \, \de_\nu \zeta^\alpha \right) \, .
\end{split}
\ee
and for the corresponding components 
\be
\begin{split}
&\delta A_1 = {\cal H} \, \zeta^0 + {\zeta^0}^\prime \, , 
\quad \delta B_1 =
\zeta^0 -\zeta^\prime \, , \quad \delta E_1 =  \zeta \, ,  \quad
\delta F_1 =  {\cal H} \, \zeta^0
\, ; \\
&\delta A_2 = {\cal H}_\beta \, \zeta^0 + {\zeta^0}^\prime \, , 
\quad \delta B_2 =
 c^2 \, \zeta^0 - \zeta^\prime \, , \quad \delta E_2 =  \zeta \, ,  \quad \delta F_2 =  {\cal H}_\o \zeta^0
\, ; \\
& \delta {\cal C}_{1/2 \, i} = {\zeta^i_T}' \, , \quad \delta {\cal
  V}_{1/2 \, i} =  {\zeta^i_T} \,  , \qquad \delta {h^{TT}}_{1/2 \, ij}  =0 \, ; \\
\end{split} 
\label{gtr}
\ee 
where 
\be
\begin{split}
&\zeta^i = \zeta^i_T + \de_i
\zeta \, , \qquad\zeta = \Delta^{-1} \de_i \zeta^i \, ,  \\[.2cm]
& {\cal H}_\beta=\frac{ (c
  \, \omega)'}{(c \, \o)} =\frac{c'}{c} +\o \, {\cal H}_\o \, .
\end{split}
\ee
In the scalar sector we have 8 fields and two independent gauge
transformations, as a result we can form 6 independent gauge invariant
scalar combinations  that we chose to be 
\be
\begin{split}
& \Psi_1= A_1 - {\cal H} \, \Xi_1   - \Xi_1^\prime \qquad \Psi_2= A_2 + c^{-2} \left(\frac{c'}{c} - {\cal H}_\o \right) \, \Xi_2   -
\frac{\Xi_2^\prime}{c^2}  \\
&\Phi_1 = F_1- {\cal H} \, \Xi_1 \, , \qquad \Phi_2 = F_2 - {\cal
  H}_\o \, \frac{\Xi_2}{c^2}  \, , \\
&{\cal E} = E_1 - E_2 \, ,  \qquad {\cal B}_1 = B_2 - c^2
B_1 +(1-c^2) \, E_1' \, ,
\end{split}
\label{sgib}
\ee
where $\Xi_{1/2} = B_{1/2} + E_{1/2}^\prime$. The following additional
gauge invariant fields will be useful to write in a compact form the
perturbed Einstein equations
\be 
\label{sgibaux}
\begin{split}
&{\cal F}_1=F_2-F_1 +\left({\cal H} - {\cal  H}_\o \right) \Xi_1 \, ,
\qquad {\cal F}_2=F_2-F_1 +\left({\cal H} -
{\cal  H}_\o \right) \Xi_2/c^2 \, , \\
&{\cal B}_2 = B_2 - c^2 B_1 +(1-c^2) \, E_2' \, ,  \\
&{\cal A}_1 = c (A_2-A_1) + \left[c \left({\cal H} -
{\cal  H}_\o \right) - c'\right] \Xi_1 \, , \\
&{\cal A}_2 = c
(A_2-A_1) + \left[c \left({\cal H} -
{\cal  H}_\o \right) - c'\right] \Xi_2/c^2 \, .
\end{split}
\ee
The fields in (\ref{sgibaux}) can be of course
expressed in terms of the ones in (\ref{sgib}).
In the matter sector, since matter is minimally coupled to the effective metric~$G_{\m\n}$, we define the gauge invariant perturbed pressure and density in the following way
\be
\delta \rho_{gi} = \delta \rho - \frac{\Xi_{\text{eff}} \, \rho'}{\left(\a\,a+\b\,\o\,c\right)^2} \, , \qquad \delta
p_{gi} = \delta p - \frac{\Xi_{\text{eff}} \, p'}{\left(\a\,a+\b\,\o\,c\right)^2}
 \,,
\ee
where
\be
\Xi_{\text{eff}} = B_{\text{eff}} + E'_{\text{eff}} \,,
\ee
and
\bea
&&B_{\text{eff}} = \a^2\,a^2\,B_1 + \frac{2\,\a\,\b\,a\,\o}{1+c}\left(c\,B_1 + B_2 \right)
+ \b^2\,\o^2\,B_2 \,,\\
&&E'_{\text{eff}}= \left(\a\,a+\b\,\o\right) \left( \a\,a\,E_1' + \b\,\o\,E_2'\right) \,.
\ena
For matter, together with pressure and density perturbation, there is
also the perturbed 4-velocity $u^\mu$ that consists of a scalar part
$v$ and a vector part~$\delta z_i$
\be
u^\mu = {\bar u}^\mu  + \delta u^\mu \, , \qquad u^\mu u^\nu G_{\mu
  \nu} = -1 \,.
\ee
The corresponding gauge invariant quantities are defined as 
\be
u_s = v + \frac{E'_{\text{eff}}}{\left(\a\,a+\b\,\o\right)^2} \, , \qquad {\delta v}_{ i} = \delta z_i + \frac{{\cal
  C}_{\text{eff} \, i}}{\left(\a\,a+\b\,\o\right)^2} \, ,  
\ee
where
\be
{\cal C}_{\text{eff} \, i} = \a^2\,a^2\,{\cal C}_{1 \, i} + \frac{2\,\a\,\b\,a\,\o}{1+c}\left(c\,{\cal C}_{1 \, i} + {\cal C}_{2 \, i} \right)
+ \b^2\,\o^2\,{\cal C}_{2 \, i} \,.
\ee
The conservation of the matter EMT ${\cal T}^{\mu \nu}$ with respect to the effective metric $G_{\m\n}$ leads to the following differential relation for vector matter perturbations
\bea
&&\delta v_i ' - \frac{1}{\left(\a\,a+\b\,\o \right)}\left[(3 \,w-1)\left(\a\,a\,\mathcal{H}+\b\,\o\,\mathcal{H}_\o\right) 
\right. \nb \\[2ex]
&& \left. \qquad +\frac{\b\,\o\left\{ \b\,\o\,c'+\a\,a\left[c'-\left(c-1\right)\left(\mathcal{H} - \mathcal{H}_\o\right)\right]\right\}}
{\left( \a\,a+\b\,\o\,c \right)}\right] \delta v_i \, =0 \, .
\label{vecc}
\ena
In the vector sector we have 4 fields and 1 gauge transformation; thus,
we can form 3 independent gauge invariant vector perturbations
\be
\begin{split}
& V_{1/2 \, i} = {\cal C}_{1/2 \, i} - {\cal V}_{1/2 \, i}^\prime \, ,
\quad \chi_i = {\cal C}_{1 \, i}-  {\cal C}_{2 \, i} \, .
\end{split}
\ee
\section{Perturbed Einstein Equations}
\subsection{Scalars}
\label{scal}
Using the definitions of the previous section we have for scalar
perturbations of $g$
\bea
&&2 \Delta  \Phi_1+6 {\cal H} \left(\Psi _1 \mathcal{H}-\Phi
  _1'\right) +a^2 \left[ m^2 \,  f_2 +  \bar \rho \, \alpha\,
  \beta  \, y^2 \, \xi \right](3 {\cal F}_1-\Delta   {\cal
  E})  \nb \\  
&&=- \frac{ \alpha \, a^2 y^3 }{2 M_{pl}^2 } \left[ \delta \rho_{\text{gi}} + \frac{\b\,\xi\,\rho'}{y_c^2}
\left( y_1\,\mathcal{B}_1 - y\,{\cal E}'\right)
 \right]\, ; \label{1tt}\\[.3cm]
&&2 \Psi _1 \mathcal{H}-2 \Phi _1' + \frac{a^2}{c+1} \left[ m^2 \, f_2 - y_3 \, \bar \rho \right]
    \mathcal{B}_1 
+ \frac{a^2 \left(1+w\right)\bar \rho\, \a\, y^3\,y_2}{ y_c} \left( u_s + \frac{\b\,\xi}{y}\, {\cal E}'\right) =0 \, ;      \label{1ts}\\[.3cm] 
&&  \left(\de_i \de_j - \delta_{ij} \Delta \right)
  \left[ a^2 \left( f_1\, m^2 - w\,\bar \rho \, d\,\xi 
    \right) \mathcal{E} - \Phi_1- \Psi_1\right]  \nb \\
&&+ \delta_{ij} \left[ 2 a^2  \left( m^2    \, f_1 - w\,\bar \rho  \, d 
          \, \xi \right) \mathcal{F}_1 
          + a^2 \left( m^2 \, f_2 -
        w\, \bar \rho\,  \a\, \beta \, y^2\, \xi \right) \mathcal{A}_1\right. \nb \\[.2cm] 
&&\left. +2 \Psi_1 \,\left(\mathcal{H}^2+2 \,\mathcal{H}'\right)
 -2 \,\Phi_1''-2\, \mathcal{H} \left(2\, \Phi _1'-\Psi
   _1'\right)\right] \nb \\
&& = \frac{ \alpha \,a^2\, y^2 \, y_c \,w}{2 M_{pl}^2}\, \delta_{ij} \,  
 \left[ \delta \rho_{\text{gi}} + \frac{\b\,\xi\,\rho'}{y_c^2}
\left( y_1\,\mathcal{B}_1 - y\,{\cal E}'\right)
 \right] \, ; \label{1ss}
\ena
where
\be
f_1 =\xi \, \left[2 \,\xi  \,\left(3\, a_3 \, c \, \xi +a_2
 \,  (c+1)\right)+a_1\right] \, , \quad f_2 =\xi \, \left(6 \,a_3 \,\xi ^2+4
\, a_2\, \xi +a_1\right) \, ,
\ee
and
\be 
\begin{split}
& y = ( \alpha + \beta \, \xi ) \, , \qquad  y_c = ( \alpha + \beta \,
c \, \xi ) \, , \qquad \bar \rho = \frac{\rho} {2
  \, M_{pl}^2} \, , \qquad d = \a \, \b \, y \, y_c \, ,\\
& y_1= \frac{2\, \a}{1+c} + \b\,\xi \, , \qquad y_2 = \a + \frac{2\,\b\,\xi\,c}{1+c} \,, \\
& y_3 = \frac{\a\,\b\, y^2\,\xi }{\left(1+c\right) y_c} \left[ c\, y  + w\left( \a\left(1+2\,c\right) +\b\,c
        \left(2+c\right)\xi\right)\right]  \,.
\end{split}
\label{def1}
\ee
For the metric $f$ we have 
\bea
&&
2\,c^2 \Delta  \Phi _2+6
   \mathcal{H}_{\omega } \left(\Psi _2   \mathcal{H}_{\omega }-\Phi
     _2'\right)  +\frac{a^2 \, c^2}{\k\,\xi^2} \left( m^2    f_2 +
   \bar \rho \, \a\, \beta  \, y^2\,\xi\right)
   \left(\Delta  \mathcal{E}-3 \,\mathcal{F}_2\right) \nb \\
&&=  -\frac{a^2\,c^2 \,\b\, y^3}{2 M_{pl}^2 \, \kappa\,\xi} \left[ \delta \rho_{\text{gi}} + \frac{\a\,\rho'}{c^2\, y_c^2}
\left( y_2\, \mathcal{B}_2 - c^2\,y\,{\cal E}'\right)
 \right]\, ; \label{2tt}\\[.3cm]
&&  2\, c \, 
   \left(\Psi _2 \mathcal{H}_{\omega }-\Phi _2'\right)- \frac{a^2}{\kappa\,\xi^2\,(1+c)}
   \left(m^2 f_2 - y_3\,\bar \rho \right) \mathcal{B}_2 \nb \\
&&+ \frac{a^2\,c^2 \left(1+w \right)\bar \rho\, \b \, y^3\,y_1 }{\kappa\,\xi\, y_c } \left(u_s - \frac{\a}{y}\,\mathcal{E}'\right) = 0\, ;
 \label{2ts}\\[.3cm]
&&
-c \, \left(\de_i \de_j - \delta_{ij} \Delta \right) \left[ 
\frac{a^2}{\k\,\xi^2} \left( f_1 \, m^2 - w\,\bar \rho\, d\,\xi \right) \mathcal{E}+ \, c   \, \left(\Phi_2+\Psi
     _2\right)\right] \nb \\
&&+\delta_{ij}\left[ \frac{2 \, a^2 \, c}{\k\,\xi^2} \left (m^2 
        f_1 -w\, \bar \rho \, d\,\xi \right)
    \mathcal{F}_2+ \frac{a^2\,c}{\k\,\xi^2}  \left( m^2 
        f_2 -w\,\bar \rho \, \a \, \b \, y^2\,\xi\right)  \mathcal{A}_2
\right. \nb \\
&& \left. +2 \left( \mathcal{H}_{\omega }^2+2\,\mathcal{H}_{\omega
  }'-2\,\frac{c'}{c}\, \mathcal{H}_{\omega }\right) \Psi _2-2 \Phi_2''+2\left(\frac{c'}{c}-2\,\mathcal{H}_{\omega
  }\right)\, 
\Phi_2'+2\,\mathcal{H}_{\omega }\,\Psi _2' \right] \nb \\
&&=  \frac{a^2
\,c}{2 \, M_{pl}^2 \,\k\,\xi}\,w \, \b \, y^2 \, 
 y_c \,  \delta_{ij} \left[ \delta \rho_{\text{gi}} + \frac{\a\,\rho'}{c^2\, y_c^2}
\left( y_2\,\mathcal{B}_2 - c^2\,y\,{\cal E}'\right)
 \right] \, . \label{2ss}
\ena
\subsection{Vectors}
\label{vect}
In the vector sector the perturbed Einstein equations are
\bea
&& \frac{\Delta V_{1 \, i}}{2 \, a^2} - 
 \frac{1}{1+c}\left( m^2 \, f_2  +\bar \rho\,\xi\,y_4 \right) \chi_i 
 - \frac{ \left(1+w\right)\bar\rho\, \a\, y^3\,y_2}{y_c}\, \delta v_i=0 \, ; \label{1vts}\\[.3cm]
&&\de_{(i} V_{1 \, j)}'  + 2 \,{\cal H}\,   \de_{(i} V_{1 \, j)}
- a^2 \left( m^2 \, f_1 - w\,\bar \rho \, d\,\xi \right)  \de_{(i}
  {\cal V}_{12 \, j)} = 0  \,
; \label{1vss} \\[.3cm]
&& \frac{\Delta V_{2 \, i}}{2 \,a^2 \, c } + \frac{1}{\kappa\, \xi^2(1+c)} \left( m^2 \, f_2
+ \bar \rho\,\xi\,y_4 \right)  \chi_i - \frac{\left(1+w \right)\bar\rho\, \b \, y^3\,y_1 }{\kappa\,\xi\, y_c } 
\, \delta v_i =0 \, ; \label{2vts}\\[.3cm]
&& 
\de_{(i} V_{2 \, j)}' +  \left[2 \left( {\cal H }  +\, \frac{\xi '}{\xi} \right)-\frac{ c'}{ c}\right] \,
  \de_{(i} V_{2 \, j)}   
+ \frac{a^2\, c}{\kappa\,\xi^2}  \left( m^2\, f_1 -w\, \bar
    \rho\,d\,\xi   \right)   \de_{(i}
{\cal V}_{12 \, j)} =0 \,;
\label{2vss}
\ena
where 
\be
{\cal V}_{12 \, i } ={\cal V}_{1\, i} -{\cal V}_{2 \, i } \, , \qquad 
{V}_{12 \, i} ={ V}_{1\, i} -{V}_{2 \, i} \, ,
\ee
and
\be
y_4= \frac{ \a\,\b\,y}{1+c}\left( 2\,y_c +\a\,c+\b\,\xi+w\,y_c\right) \,.
\ee
Notice that ${V}_{12 \,i} = \chi_i -{\cal V}_{12 \, i }' $. 
\end{appendix}


\begin{thebibliography}{99} 

\bibitem{megarevs}
   T.~Clifton, P.~G.~Ferreira, A.~Padilla and C.~Skordis,
  Phys.\ Rept.\  {\bf 513}, 1 (2012)
  [arXiv:1106.2476 [astro-ph.CO]].
  
  A.~Joyce, B.~Jain, J.~Khoury and M.~Trodden,
  arXiv:1407.0059 [astro-ph.CO].


\bibitem{deRham}
 C.~de Rham,
   Living Rev.\ Rel.\  {\bf 17}, 7 (2014)
   [arXiv:1401.4173 [hep-th]].


\bibitem{Damour} 
  T.~Damour and I.~I.~Kogan,
  Phys.\ Rev.\ D {\bf 66}, 104024 (2002)
  [hep-th/0206042].


\bibitem{Berezhiani} 
  Z.~Berezhiani, D.~Comelli, F.~Nesti and L.~Pilo,
  Phys.\ Rev.\ Lett.\  {\bf 99}, 131101 (2007)
  [hep-th/0703264 [HEP-TH]].

\bibitem{Comelli} 
  D.~Comelli, M.~Crisostomi, F.~Nesti and L.~Pilo,
  Phys.\ Rev.\ D {\bf 85}, 024044 (2012)
  [arXiv:1110.4967 [hep-th]].


\bibitem{Hassan:2011zd}
  S.~F.~Hassan and R.~A.~Rosen,
  JHEP {\bf 1202} (2012) 126
  [arXiv:1109.3515 [hep-th]].


\bibitem{FRWfroz}
G. D'Amico, C. de Rham, S. Dubovsky, G. Gabadadze, D. Pirtskhalava,
A.J. Tolley,
Phys.\ Rev.\ D {\bf 84} (2011) 124046
[arXiv:1108.5231 [hep-th]].

\bibitem{open} 
  A.E.~Gumrukcuoglu, C.~Lin and S.~Mukohyama,
  \emph{JCAP} {\bf 1111}, 030 (2011)
  [arXiv:1109.3845 [hep-th]].

\bibitem{tasinato} 
N.~Khosravi, G.~Niz, K.~Koyama and G.~Tasinato,
  JCAP {\bf 1308}, 044 (2013)
  [arXiv:1305.4950 [hep-th]].
  
\bibitem{defelice} 
  A.~De Felice, A.~E.~Gumrukcuoglu and S.~Mukohyama,
  Phys.\ Rev.\ Lett.\  {\bf 109}, 171101 (2012)
  [arXiv:1206.2080 [hep-th]].

  A.~De Felice, A.E.~Gumrukcuoglu, C.~Lin and S.~Mukohyama,
  \emph{JCAP} {\bf 1305} (2013) 035,  [arXiv:1303.4154 [hep-th]].


\bibitem{Comelli:2013tja} 
  D.~Comelli, F.~Nesti and L.~Pilo,
  JCAP {\bf 1405}, 036 (2014)
  [arXiv:1307.8329 [hep-th]].

\bibitem{lang}
  D.~Langlois, S.~Mukohyama, R.~Namba and A.~Naruko, Class.\ Quant.\ Grav.\  {\bf 31} (2014) 175003
  [arXiv:1405.0358 [hep-th]].


\bibitem{Rubakov} 
  V.A.~Rubakov, 
  [arXiv:hep-th/0407104]. 

\bibitem{dub} 
   S.L.~Dubovsky, 
 \emph{JHEP} {\bf 0410}, 076 (2004) [hep-th/0409124].

\bibitem{uscan} 
D.~Comelli, M.~Crisostomi, F.~Nesti and L.~Pilo,
  Phys.\ Rev.\ D {\bf 86}, 101502 (2012).
  [arXiv:1204.1027 [hep-th]].

\bibitem{uslong} 
  D.~Comelli, F.~Nesti and L.~Pilo,
  JHEP {\bf 07} 161 (2013)  
  [arXiv:1305.0236 [hep-th]].

\bibitem{Comelli:2014xga} 
  D.~Comelli, F.~Nesti and L.~Pilo,
  JCAP {\bf 1411}, no. 11, 018 (2014)
  [arXiv:1407.4991 [hep-th]].


\bibitem{cosm-CCNP} 
  D.~Comelli, M.~Crisostomi, F. Nesti, L.~Pilo, 
  \emph{JHEP} {\bf 1203} (2012) 067, 
  Erratum-ibid.\ {\bf 1206} (2012) 020   [arXiv:1111.1983 [hep-th]].
 
\bibitem{bi-GF}
  M.~von Strauss, A.~Schmidt-May, J.~Enander, E.~Mortsell and S.F.~Hassan, 
  \emph{JCAP} {\bf 1203} (2012) 042 [arXiv:1111.1655 [gr-qc]].
  
  M.S.~Volkov, 
  \emph{JHEP} {\bf 1201} (2012) 035 [arXiv:1110.6153 [hep-th]].
  
  Y.~Akrami, T.~S.~Koivisto and M.~Sandstad,
  JHEP {\bf 1303}, 099 (2013)
  [arXiv:1209.0457 [astro-ph.CO]].
  

\bibitem{cosmpert} 
  D.~Comelli, M.~Crisostomi and L.~Pilo,
  JHEP {\bf 1206}, 085 (2012)
  [arXiv:1202.1986 [hep-th]].

\bibitem{pertslast} 
  D.~Comelli, M.~Crisostomi and L.~Pilo,
  Phys.\ Rev.\ D {\bf 90}, 084003 (2014)
  [arXiv:1403.5679 [hep-th]].


\bibitem{Lagos:2014lca} 
  M.~Lagos and P.~G.~Ferreira,
  [arXiv:1410.0207 [gr-qc]].
  
\bibitem{Cusin:2014psa}
  G.~Cusin, R.~Durrer, P.~Guarato and M.~Motta,
  [arXiv:1412.5979 [astro-ph.CO]].

\bibitem{deRham:2014naa} 
  C.~de Rham, L.~Heisenberg and R.~H.~Ribeiro,
  [arXiv:1408.1678 [hep-th]].
  
\bibitem{Heisenberg:2014rka} 
  L.~Heisenberg,
  arXiv:1410.4239 [hep-th].
  
\bibitem{Tamanini:2013xia} 
  N.~Tamanini, E.~N.~Saridakis and T.~S.~Koivisto,
  JCAP {\bf 1402}, 015 (2014)
  [arXiv:1307.5984 [hep-th]].
  
\bibitem{Noller:2014sta} 
  J.~Noller and S.~Melville,
  arXiv:1408.5131 [hep-th].

\bibitem{Solomon:2014iwa}
  A.~R.~Solomon, J.~Enander, Y.~Akrami, T.~S.~Koivisto, F.~Konnig and E.~Mortsell,
  [arXiv:1409.8300 [astro-ph.CO]].

\bibitem{Enanderbef} 
  J.~Enander, A.~R.~Solomon, Y.~Akrami and E.~Mortsell,
  JCAP {\bf 01}, 006 (2015)
  [arXiv:1409.2860 [astro-ph.CO]].

\bibitem{deRham-ghost} 
  C.~de Rham, L.~Heisenberg and R.~H.~Ribeiro,
  [arXiv:1409.3834 [hep-th]].

\bibitem{Yamashita:2014fga} 
  Y.~Yamashita, A.~De Felice and T.~Tanaka,
  [arXiv:1408.0487 [hep-th]].
  
  
\bibitem{Gumrukcuoglu:2014xba} 
 A.~E.~Gumrukcuoglu, L.~Heisenberg and S.~Mukohyama,
  [arXiv:1409.7260 [hep-th]].


\bibitem{strau}
General Relativity, Second Edition; Chapter 3.5. N. Straumann;  Springer.

\bibitem{giulini}
D.~Giulini,
 Lect.\ Notes Phys.\  {\bf 721}, 105 (2007)
 [gr-qc/0603087].
 
  \bibitem{Gabadadze:2011}   
  C.~de Rham, G.~Gabadadze, A.J.~Tolley, 
    {Phys.\ Rev.\ Lett.}\  {\bf 106}, 231101 (2011)
  [arXiv:1011.1232 [hep-th]]. 
 
 \bibitem{Hassan:2011vm} 
  S.~F.~Hassan and R.~A.~Rosen,
  JHEP {\bf 1107}, 009 (2011)
  [arXiv:1103.6055 [hep-th]].

\bibitem{HR}
S.~F.~Hassan and R.~A.~Rosen,
  Phys.\ Rev.\ Lett.\  {\bf 108} (2012) 041101.
  [arXiv:1106.3344 [hep-th]].
  
\bibitem{vain}
E.~Babichev and M.~Crisostomi,
  Phys.\ Rev.\ D {\bf 88}, 084002 (2013)
  [arXiv:1307.3640];
  

\bibitem{Schmidt-May:2014xla} 
  A.~Schmidt-May,
  arXiv:1409.3146 [gr-qc].

\bibitem{deRham:2014tga}
  C.~de Rham, A.~Matas, N.~Ondo and A.~J.~Tolley,
  [arXiv:1410.5422 [hep-th]].


\bibitem{Tasinato:2014eka}
  G.~Tasinato,
  JHEP {\bf 1404} (2014) 067
  [arXiv:1402.6450 [hep-th]].

\bibitem{Heisenberg:2014rta}
  L.~Heisenberg,
  JCAP {\bf 1405} (2014) 015
  [arXiv:1402.7026 [hep-th]].
  
\bibitem{Konnig:2014xva}  
  A.~R.~Solomon, Y.~Akrami and T.~S.~Koivisto,
  JCAP {\bf 1410}, 066 (2014)
  [arXiv:1404.4061 [astro-ph.CO]].

  F.~Koennig, Y.~Akrami, L.~Amendola, M.~Motta and A.~R.~Solomon,
  Phys.\ Rev.\ D {\bf 90}, no. 12, 124014 (2014)
  [arXiv:1407.4331 [astro-ph.CO]].
 



\end{thebibliography}
\end{document}